# Study of two-subband population in Fe-doped Al$_x$Ga$_{1-x}$N/GaN heterostructures by persistent photoconductivity effect

Ikai Lo,[a)] J.K. Tsai, M.H. Gau, Y.L. Chen, Z.J. Chang, W.T. Wang, J.C. Chiang, and K.R. Wang[b)]

Department of Physics, Center for Nanoscience and Nanotechnology, National Sun Yat-Sen University, Kaohsiung, Taiwan, Republic of China

Chun-Nan Chen

Department of Electronic Engineering, Far-East College, Hsin-Shih Town, Tainan, Taiwan, Republic of China

T. Aggerstam

KTH Royal Institute of Technology, Electrum 229, SE-164 40 Kista, Sweden

Abstract

The electronic properties of Fe-doped Al$_{0.31}$Ga$_{0.69}$N/GaN heterostructures have been studied by Shubnikov-de Haas measurement.  Two subbands of the two-dimensional electron gas in the hetero-interface were populated.  After the low temperature illumination, the electron density increases from 11.99 x 10$^{12}$ cm$^{-2}$ to 13.40 x 10$^{12}$ cm$^{-2}$ for the first subband and from 0.66 x 10$^{12}$ cm$^{-2}$ to 0.94 x 10$^{12}$ cm$^{-2}$ for the second subband.  The persistent photoconductivity effect (~13% increase) is mostly attributed to the Fe-related deep-donor level in GaN layer.  The second subband starts to populate when the first subband is filled at a density $n_1$ = 9.40 x 10$^{12}$ cm$^{-2}$.  We calculate the energy separation between the first and second subbands to be 105 meV.

III-nitride heterostructures have been extensively studied due to the applications in high-electron-mobility transistors[1] and high-bright blue laser diodes.[2] The donor-related deep level is one of the important issues in the performance for both electronic and optical device applications. In the earlier years, the lattice mismatch between GaN and substrate (e.g., sapphire) gave rise to a high density of threading dislocations as grown by metal-organic chemical vapor deposition (MOCVD).[3,4] By an appropriate growth technique (such as using a template before the growth of GaN epilayer), the density of threading dislocations can be apparently reduced,[5] and hence a high-quality GaN-based heterostructure is achievable. However, in addition to the threading dislocations, the impurities and native defects (e.g., N vacancy,[6] O-residual donor,[7] and Si-residual donor[8]) are generated during the fabrication of Si-doped $Al_xGa_{1-x}N$/GaN heterostructures and result in a high density of $n$-type carriers in the sample. The high carrier concentration indicates the presence of high density of defects. Our previous study showed that the increment of carrier concentration in Si-doped $Al_xGa_{1-x}N$/GaN heterostructures measured by persistent photoconductivity (PPC) effect was less than 2.5 %.[9] The PPC effect is a common property of $Al_xGa_{1-x}As$ in which the carrier concentration is increased by illuminating the sample at low temperature (e.g., < 77K) and the increment of carrier concentration is persistent even after the light is removed. The carrier concentration increases due to the photo-excited electrons transferred from the deep-level donors (e.g., DX-centers in $Al_xGa_{1-x}As$)[10] to the conduction band when the light is turned on. The local potential barrier around the deep-level donors prevents the recombination of the electrons and the ionized deep-level donors after the light is off at low temperature and hence the increment of carrier concentration persists. The PPC effect in $Al_xGa_{1-x}As$ is induced by the metastable DX-centers.[10] However, when a heterostructure (or quantum well) is made of semiconductor compounds, the PPC effect is no longer limited to the presence of DX-centers. The other deep-level donors in the heterostructure are able to produce a PPC effect as long as the deep donor level is below Fermi energy.[11] Recently, a high-electron-mobility transistor was made of $Al_xGa_{1-x}N$/GaN heterostructure with Fe-doped GaN layer instead of the traditional Si-doped $Al_xGa_{1-x}N$ layer.[12] It was shown that the electronic properties were enhanced by furnace annealing, indicating that the structural defects are expected in this Fe-doped $Al_xGa_{1-x}N$/GaN high-electron-mobility transistor.  In this paper, we presented the results of Shubnikov-de Haas (SdH) measurement on a similar high-electron-mobility $Al_xGa_{1-x}N$/GaN heterostructure and found a large PPC effect in the samples.  Using the PPC effect as a tool to vary the carrier concentrations of the lowest two subbands in two-dimensional electron gas (2DEG), we obtain the energy separation between the first and second subband minima ($\Delta E_{12} = E_2 - E_1$).  In general, the energy separation can be obtained either from the theoretical calculation or from the experimental two-subband-populated result fitting to equation, $\Delta E_{12} = \pi\hbar^2 n_1/m_1^* - \pi\hbar^2 n_2/m_2^*$, where $n_i$ and $m_i^*$ are the carrier concentrations and effective mass of $i$'th subbands, respectively. The carrier concentrations can be obtained, experimentally, from the SdH frequency, $n_i = 2ef_i/h$, where $f_i$ is the SdH frequency of $i$'th subband and the reduced Planck's constant $\hbar = h/2\pi$.  In the literature, it is always used the same mass for the first and second subband bands to calculate the energy separation.  However, the effective mass for the two subbands is, in fact, not exactly the same due to the non-parabolic effect.  The second subband effective mass can be experimentally determined by cyclotron resonance (CR) measurement.  For instance, the effective masses of the first and second subbands are $m_1^*=0.060m_0$ and $m_2^*=0.049m_0$, respectively, for the 2DEG of AlInAs/InGaAs heterostructure.[13]  Since the second subband effective mass ($m_2^*$) for AlGaN/GaN is not available up to date, the energy separation obtained from the onset of the second subband population in this study becomes the direct measurement without the assumption of the same effective mass ($m_1^* = m_2^*$), in which the information of the second subband mass is not necessary.[14]

The $Al_xGa_{1-x}N$/GaN samples were grown by metal-organic chemical vapor deposition on a *c*-plane sapphire substrate with ferrocene ($Cp_2Fe$) as an iron source. Trimethylgallium (TMG) and trimethylaluminum (TMAl) and ammonia were used as precursors.  Above the *c*-plane sapphire substrate, the sample structure consists of a 100-nm-thick low-temperature GaN buffer layer, 800-nm-thick Fe-doped GaN layer, unintentionally doped 1.6-μm-thick GaN layer, and unintentionally doped 20-nm-thick $Al_xGa_{1-x}N$ barrier layer.  The similar sample structure has been fabricated for a high-electron-mobility transistor.[12] A Hall-bar-shaped $Al_xGa_{1-x}N$/GaN sample (x = 0.31) with indium ohmic contacts was annealed at 350 $^o$C for 5 minutes under $N_2$ forming gas. The mobility of the sample (determined from Hall measurement at the temperature of 0.38 K) is 6.4 x $10^3$ $cm^2$/Vs, and the corresponding carrier concentration 12.79 x $10^{12}$ $cm^{-2}$. After an extensive illumination (>10561 seconds), the mobility decreases to 5.4 x $10^3$ $cm^2$/Vs, but the carrier concentration increases to 14.70 x $10^{12}$ $cm^{-2}$. The PPC conditions were provided by illuminating the sample at T ~ 0.38 K for different time periods using a blue light-emitting diode with 472-nm peak wavelength. After the illumination we performed the Shubnikov-de Haas measurement on the sample for the magnetic field (*B*) swept from 0.5 T to 12 T at temperatures about 0.38 K. Because the oscillation part of Subnikov-de Haas (SdH) effect is a function of cosine, e.g., $\cos(2\pi f_{SdH}/B)$, the SdH data (2048 points) were taken by equal spacing in 1/*B* for the purpose of fast Fourier transformation (FFT). In our case (2048 points for the magnetic field from 0.5 T to 12 T), the resolution of SdH frequency for the FFT spectrum is equal to 0.26 T.

The magnetoresistance ($R_{XX}$) versus magnetic field for different illumination times was plotted in Figure 1, where the zeroes of y-axes have been offset for comparison. The data at the top is the result of non-illuminated SdH measurement, and the data at the bottom is the result obtained after illuminating the sample for 10561 seconds. There are two well-pronounced SdH oscillations detected. The two fundamental oscillations ($f_1$ and $f_2$) correspond to the SdH oscillations for the carrier concentrations in the first- and second-lowest subbands ($E_1$ and $E_2$) of the 2DEG confined in the hetero-interface. The fast Fourier transformations of these SdH oscillations were shown in Fig. 2. It is found that the amplitudes of the two SdH oscillations ($f_1$ and $f_2$) are both increased slightly with the illumination time and reach the maximum values at 421-second illumination (marked arrows in Figs. 1 and 2). The amplitudes of both oscillations decrease when the illumination time is longer than 421 seconds and reach the minimum values at 3171-second illumination (marked arrows in Figs. 1 and 2), where the amplitude of $f_1$-SdH oscillation is even merged with the secondary peaks beside. After extensive illumination, both amplitudes increase again but the frequencies nearly reach saturation. In addition to the fundamental oscillations ($f_1$ and $f_2$), two second harmonics which have the frequencies double of the fundamental ones (i.e., the peaks at the frequencies about $2f_1$ and $2f_2$) are also observed in Fig. 2. The second harmonics are always presented in the FFT spectrum, particularly, when the SdH oscillation is accompanied with noise. Moreover, there are several peaks on the shoulders of $f_1$-SdH peak at the frequencies about ($f_1 \pm f_2$), which might be due to the magneto-intersubband resonant scattering.[15-17] For a high mobility 2DEG with two subbands populated, the oscillatory magnetoresistance will contain four components: two fundamental SdH oscillations (with SdH frequencies, $f_1$ and $f_2$) and two magneto-intersubband resonant scattering (MIS) terms (with MIS frequencies, $f_1 \pm f_2$),. Since the amplitudes of the SdH oscillations are the first-order terms of ($1/g_0$) and those of MIS oscillations are the second-order terms of ($1/g_0^2$), the amplitudes of the MIS components at ($f_1 \pm f_2$) are always smaller than those of SdH components at low temperatures, where $g_0$ is the density of state for a single subband in zero field.[15] In order to confirm the two SdH oscillations, we applied a nonlinear curve-fitting technique to the original data in Fig. 1 to decouple the SdH components from the others. After the removal of the monotonous background, the oscillatory resistivity $\rho_{osc}(B)$ was fitted to the superposition of two independent SdH cosine functions,[18]

$$\rho_{osc}(B) = \sum_{i=1}^{2} \rho_i \frac{(\mu_i B)^2}{1+(\mu_i B)^2} \exp(\frac{-\pi}{\mu_i B}) \frac{1/\xi_i B}{\sinh(1/\xi_i B)} \cos(2\pi f_i / B + \phi_i),$$

where $\rho_i$ is a constant proportional to the zero-field resistivity, $\mu_i = e\tau_i/m^*$, $\tau_i$ is the quantum lifetime of the carrier, $\xi_i = \hbar e/2\pi^2 k_B T m^*$, $f_i$ and $\phi_i$ are the SdH frequency and phase constant of the $i$th subband. Figure 3 are the results for the data of SdH measurement after 10561-second illumination (the bottom one in Fig. 1). The original data (i) is plotted in Fig. 3a and its FFT spectrum is shown in Fig. 3b. The result of nonlinear curve fit (ii) is plotted in Fig. 3a and the FFT spectrum of the nonlinear curve-fitting (nlcf) data is shown in Fig. 3b. The FFT spectrum of the nlcf data shows two perfect SdH oscillations with frequencies of $f_1 = 276.4$ T and $f_2 = 19.7$ T, and phase constants $\phi_1 = 137°$ and $\phi_2 = 134°$. The SdH components are decoupled by subtracting the nlcf data from the original data. The residual data is shown in Fig. 3a (iii), and again we calculate the FFT spectrum of the residual data and show in Fig. 3b. It is shown that the second harmonic terms (at frequencies about $2f_1$ and $2f_2$) and the components at ($f_1 \pm f_2$) are left behind after decoupling of SdH terms. In addition, there exists a peak at frequency of 238.2 T, and the three-peak patterns in FFT spectrum have been observed in Tsubaki's paper as well, in which they attributed the third peak to the spin splitting of the first subband.[19] We do not eliminate the possibilities that it might be produced due to the presence of spin splitting.

The carrier concentrations of the two subbands were calculated from the SdH frequencies, $n_i = 2ef_i/h$, for the different illumination times. The carrier concentrations in the first and second subbands before illumination are $n_1 = 11.99 \times 10^{12}$ cm$^{-2}$, and $n_2 = 0.66 \times 10^{12}$ cm$^{-2}$. They are in good agreement with the Hall measurement ($n_H = 12.79 \times 10^{12}$ cm$^{-2}$). After the extensive illumination (e.g., 10561 seconds), the carrier concentrations increase to $n_1 = 13.40 \times 10^{12}$ cm$^{-2}$, and $n_2 = 0.94 \times 10^{12}$ cm$^{-2}$, and the total carrier concentration $n_T$ increases from $12.65 \times 10^{12}$ cm$^{-2}$ to $14.34 \times 10^{12}$ cm$^{-2}$. The total carrier concentration increases by 13% due to the extensive illumination. In our previous study, the increment produced in persistent photoconductivity was less than 2.5% for Si-doped Al$_x$Ga$_{1-x}$N/GaN heterostructures grown by MOCVD.[9] It indicates that the common defects or dislocations in MOCVD grown Al$_x$Ga$_{1-x}$N/GaN will not produce a large PPC effect. The samples used here have the similar simple structures, but different impurity dopants: Fe-doped in GaN layer. Because the electron mobility of the Fe-doped sample ($\mu_H = 6.4 \times 10^3$ cm$^2$/Vs) is higher than that of Si-doped Al$_{0.31}$Ga$_{0.69}$N/GaN heterostructure used in our previous study ($\mu_H = 4.0 \times 10^3$ cm$^2$/Vs and $n_H = 12.4 \times 10^{12}$ cm$^{-2}$),[9] the defects or dislocations should be less as compared to the Si-doped Al$_{0.31}$Ga$_{0.69}$N/GaN. Moreover, the defects or dislocations generated in the Si-doped Al$_{0.31}$Ga$_{0.69}$N/GaN are more likely to form shallow donors which contribute to the high density of *n*-type carriers and thus can not produce a large PPC effect. Therefore the large PPC effect in the sample is contributed from the Fe-related deep-level donors, due to the higher localization and large lattice relaxation of Fe impurities as compared to Si impurities. The SdH frequencies versus illumination time are plotted in Fig. 4a. The discontinuity of $f_1$ occurs at 3171-second illumination before saturation. It is about the beginning of the growth of three-peak pattern in the FFT spectra (Fig. 2). As we mentioned before, there are two possibilities to create the three-peak pattern in SdH measurement: the MIS term and the spin-splitting first subband. Further evidence is needed to clarify the third peak. The carrier concentrations before saturation ($n_1$ and $n_2$) are plotted against the total carrier concentration ($n_T = n_1 + n_2$) in Fig. 4b. The increase rates of the carrier concentrations are obtained from the plot: $n_1 = 1.926 + 0.795 n_T$ and $n_2 = -1.926 + 0.205 n_T$. The onset of the second-subband population occurs at $n_1 = 9.40 \times 10^{12}$ cm$^{-2}$. The energy separation ($E_2-E_1$) can be determined from the energy level that the second subband starts to populate (the energy level, $E_F = \pi \hbar^2 n_1/m^*$, at $n_1 = 9.40 \times 10^{12}$ cm$^{-2}$). We use the effective mass for the first subband $m^* = 0.215 m_0$ and calculate the energy separation $E_2-E_1 = 105$ meV. The result is consistent with the theoretical value calculated by Garrido et al. ($E_2-E_1 = 112$ meV).[20]

In order to evaluate the subband levels of 2DEG versus the Fe-related deep-donor levels, we calculated the band potential profile (the solid line) and electron distribution (the dotted line) of Al$_{0.31}$Ga$_{0.69}$N/GaN heterostructure by a self-consistent solution of Poisson-Schrodinger equation[21] and showed in Fig. 5, where we set $x = 0.31$, the conduction band offset $\Delta E_C = 0.55$ eV and the carrier concentration of bound states at the hetero-interface $n_{2D} = 12.7 \times 10^{12}$ cm$^{-2}$ (the case of our sample). The material parameters for the piezoelectric coefficients and spontaneous polarization in Table I of ref. [20] were used. We obtained that the energy separation between the first and second subbands is 105.6 meV and the polarization electric field at the interface is 1.73 MV/cm. It is the high polarization electric field to confine the carriers of first and second subbands at the hetero-interface and hence induces a high carrier concentration and large energy separation. For the third subband, the electron wave function is more extensive to the flat conduction band. It is noted that the Fe-doped GaN layer is located at 1.6 μm far away from the hetero-interface in our sample, but a Si-doped Al$_x$Ga$_{1-x}$N barrier is 3 nm (the spacer) apart from the interface in ref. [20]. Recently, the deep-level electron traps in Fe-doped GaN have been studied.[22-24] Polyakov et al. found that the thermal activation energy of the deep-level donor is about 0.5 eV, and the photo-activated deep electron traps is at the level ($E_C$ − 0.9 eV) and hole traps at ($E_V$ + 0.9 eV), which provided promising insulating buffers for $Al_xGa_{1-x}N$/GaN heterostructures.[22] Those deep donor levels are much lower than the Fermi level and hence are occupied at low temperature (e.g., T ~ 0.38 K) before illumination. After the illumination, the electrons were transferred from the deep-level donors to the 2DEG at hetero-interface. The carrier concentration in the bound states ($n_{2D}$) increases. The calculated results are consistent with our conclusion that the large increment of carriers (~13%) by PPC effect is mostly ascribed to the Fe-related deep-donor level in GaN layer.

In conclusion, we have studied the electronic properties of Fe-doped $Al_xGa_{1-x}N$/GaN heterostructures by Shubnikov-de Haas measurement. Two-subband-populated 2DEG was detected. After the illumination of the sample at low temperature, the total carrier concentration increased from $12.65 \times 10^{12}$ cm$^{-2}$ to $14.34 \times 10^{12}$ cm$^{-2}$. A large persistent photoconductivity effect (~13% increase) is observed and we attributed the PPC effect to the Fe-related deep-donor level in GaN layer. From the increase rates of the carrier concentrations for the two subbands, we calculated the energy separation between the first and second subbands to be 105 meV.

The authors are grateful to X. Y. Liang, J. B. Yu, and X. Z. Chang for their assistance. This project is supported in part by National Research Council of Taiwan and NRC Core Facilities Laboratory for Nanoscience and Nanotechnology in Kaohsiung-Pingtung Area.

**Figure captions:**

Fig. 1. The magnetoresistance $R_{XX}$ versus magnetic field for different illumination times.

Fig. 2. Fast Fourier transformation of the magnetoresistance $R_{XX}$ for different illumination times.

Fig. 3. (a) The magnetoresistance $R_{XX}$ of (i) original data, (ii) non-linear curve fitting data, and (iii) the residual data. (b) FFT spectra of (i) original data, (ii) non-linear curve fitting data, and (iii) the residual data.

Fig. 4. (a) The SdH frequencies against illumination time for the first and second subbands. (b) The carrier concentrations of the first and second subbands versus total carrier concentration determined from the SdH frequencies.

Fig. 5. The calculated band potential profile and electron distribution in $Al_xGa_{1-x}N/GaN$ heterostructure with x = 0.31. The Fe-related deep-level donors reside at the remote Fe-doped GaN layer (1.8 μm away from the hetero-interface).

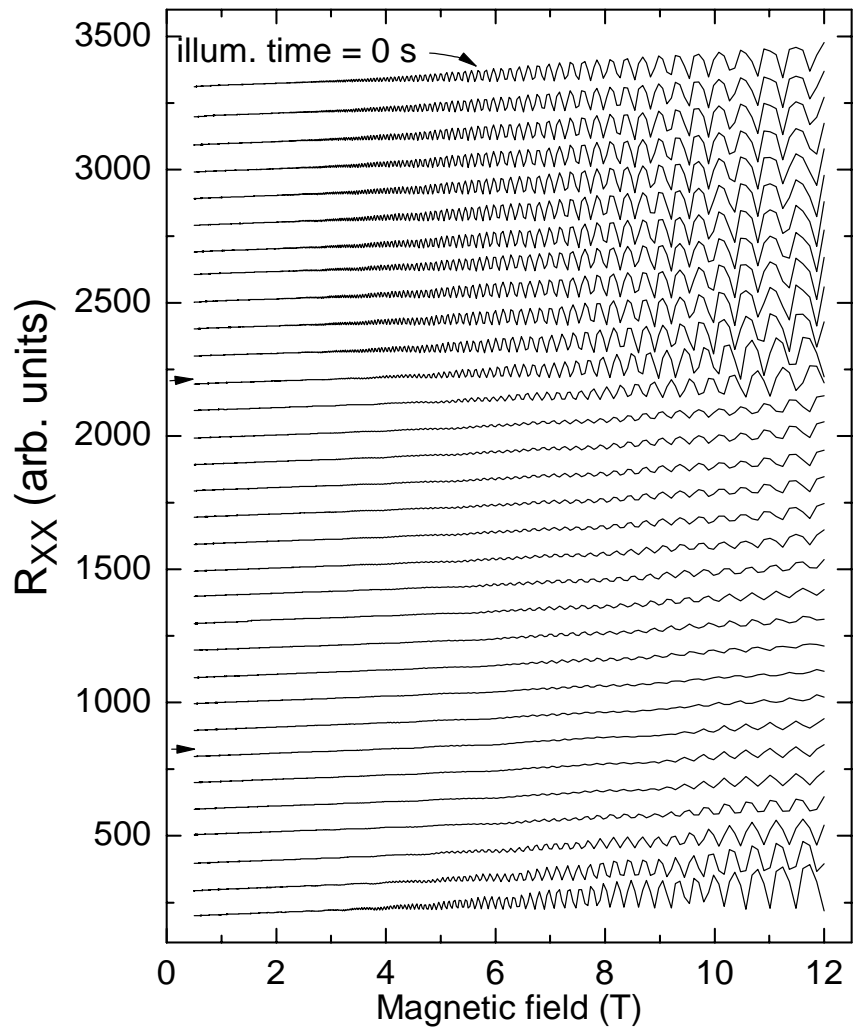

Fig. 1.

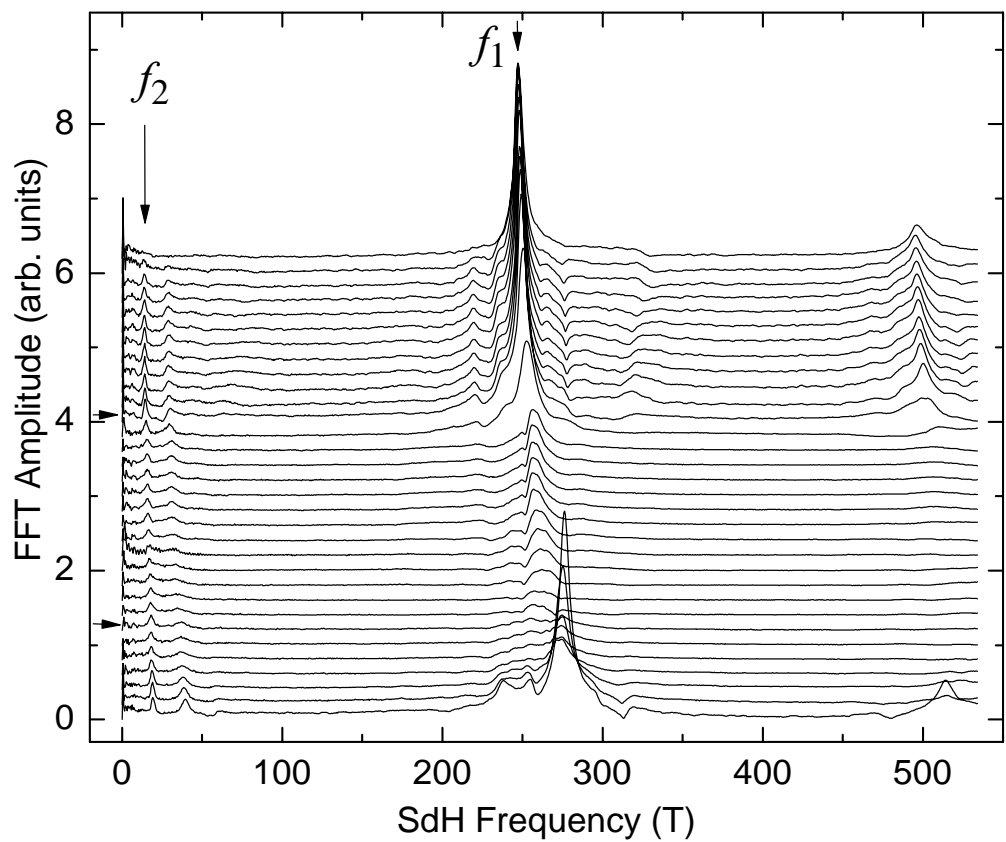

Fig. 2.

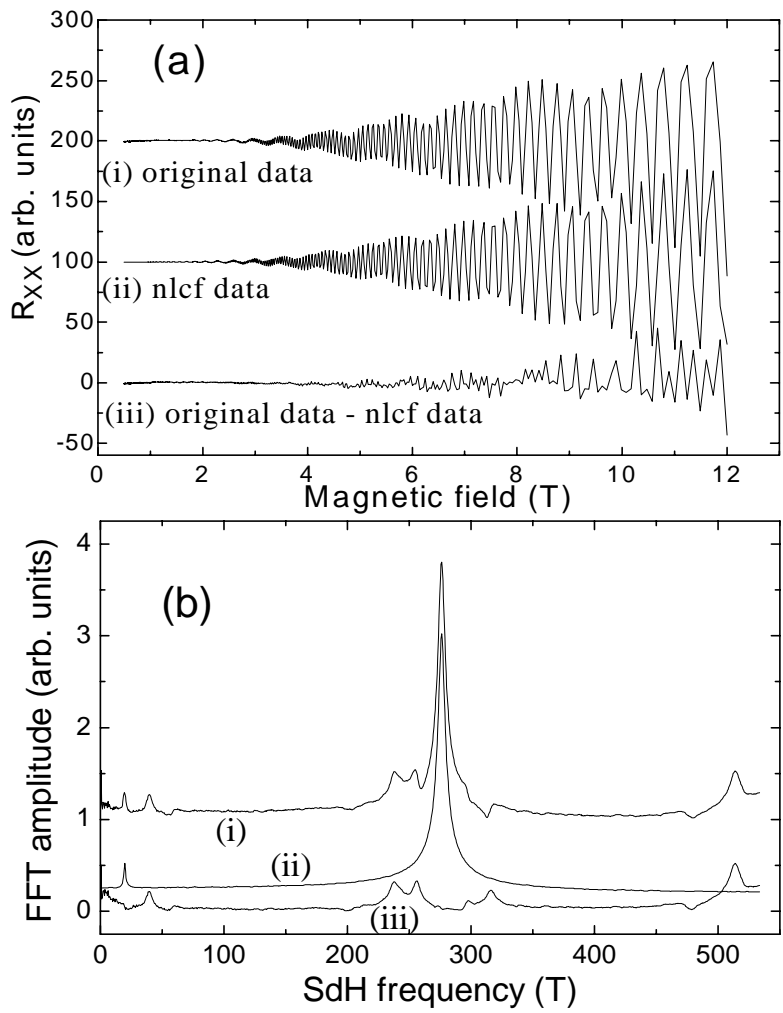

Fig. 3.

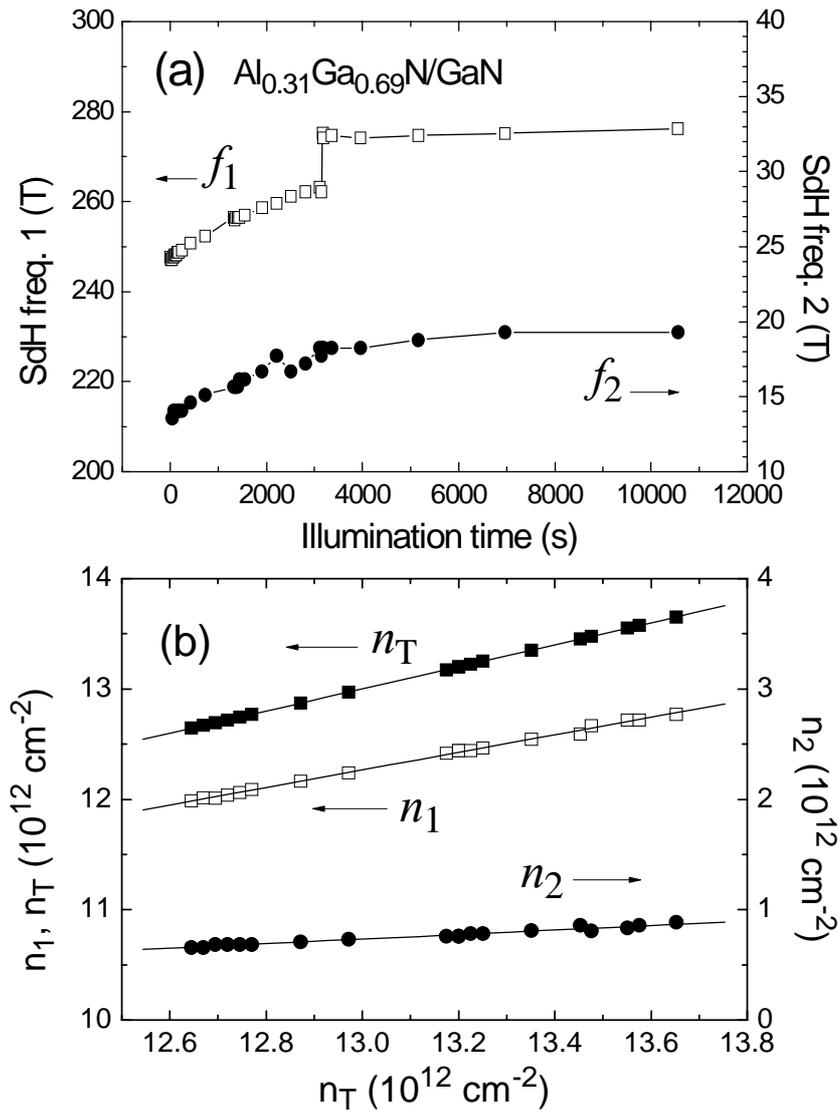

Fig. 4.

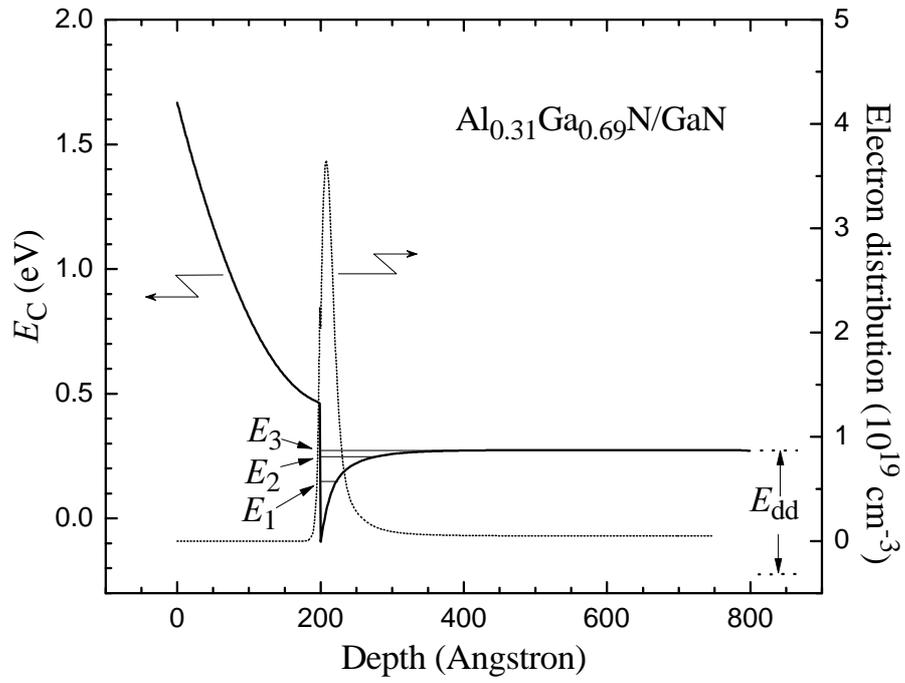

Fig. 5.